\documentstyle{article}
\input tcilatex

\begin{document}

\begin{center}
{\bf UV (}$h\nu =8.43${\bf \ eV) PHOTOELECTRON SPECTROSCOPY OF POROUS
SILICON NEAR FERMI LEVEL}
\end{center}

\begin{quote}
A.M. Aprelev, A.A. Lisachenko 

{\it Institute of Physics, St. Petersburg University, 198904, St.
Petersburg, Russia}

R. Laiho, A. Pavlov and Y. Pavlova 

{\it Wihuri Physical Laboratory, University of Turku, FIN 20014 Turku,
Finland}
\end{quote}

\begin{quotation}
\vspace{1.0in}Electronic spectra of porous Si have been investigated in the
region $\approx $4 eV below the Fermi level with specimens subjected to{\it %
\ in situ} oxygenation and thermal treatments. The measurements were made
with a UHV photoelectron spectrometer using ''soft'' energy ($h\nu =8.43$
eV) excitation of the photoemission. The significance of DOS to the
photoluminescence and its degradation in porous Si is discussed. Fine
structure of the photoelectron spectra is found from specimens heated in
oxygen at 600 K.
\end{quotation}

\newpage\ Despite of intensive studies during the past few years the
mechanisms of the photoluminescence and its degradation in porous Si are
partly unclear. Two main mechanisms proposed for explanation of the
photoluminescence are (i) quantum confinement [1] and (ii) chemical
modification of the surface [2,3]. The quantum-dot model has been applied to
porous Si by assuming escape of carriers through oxide barriers surrounding
exciton confinement regions [4]. It is obvious that changes of the atomic
structure of the surface will influence on the effective electronic
structure of silicon nanocrystals and the carrier dynamics. As shown by
secondary ion mass spectroscopy (SIMS) investigations exposure of porous Si
to ambient or various other gases results in measurable changes of the
chemical composition of the surface [5]. Therefore detailed information
about electrophysical parameters (work function $\varphi _T$, dipole
component $\delta $ of $\varphi _T$, band bending V$_S$ and spectra of the
density of occupied states (DOS)) and their change during reaction of the
surface with gaseous molecules is desirable.

In the present paper we report an investigation into electrophysical
properties and electronic states near the Fermi level (E$_F$) of porous Si
using photoelectron spectroscopy with low energy excitation. Previous
measurements of photoelectron spectra in this material have been made with
syncrotron [6,7] or x-ray [8] radiation and deal mainly with Si 2{\it p}
core-level and valence band features or by using a He-I resonance lamp for
excitation [9].

Porous Si samples with typical size of 1 cm$^2$ were prepared by
electrochemical anodization of a p-type 5 cm Si(100) wafer for 5 min at
current density of 25 mA/cm$^2$ in a 1:1:2 mixture of HF, water and ethanol.
The distribution of photoluminescence was unhomogeneous over the surface of
the samples: there were large domains with strong red photoluminescence
separated from others showing weak yellowish luminescence. An UV
photoelectron spectrometer [10] with ultrasoft excitation by Xe line ($h\nu
=8.43$eV) and an extremely low (5$\cdot $10$^{-4}$) background radiation for
measuring weak signals near the Fermi level was used for the measurements.
The excitation light was condensed inside an elliptical spot of 2$\times $3
mm on the surface of the sample. With this equipment it is possible to
determine features of DOS spectra, which are masked under excitation made
with more energetic He-I and He-II lines or synchrotron radiation. The
spectrometer has the following parameters: the resolution, determined from
the Au spectrum, is better than 60 meV and the accuracy referred to the
Fermi level is 10 meV. The base pressure of the spectrometer is not worse
than 5$\cdot $10$^{-10}$ Torr. The sample was cleaned in situ by heating in
vacuum or in an oxygen flow. High-purity oxygen (99.99\%) was used in the
experiments. Also UV-irradiation was used for in situ desorption of gas
contaminants. A special feature is that evolution of the photoelectron
spectra during thermo- and photoactivated desorption (adsorption) of oxygen
was monitored. The position of E$_F$ was determined through calibration with
the photoelectron spectrum of an Au foil cleaned in ultra high vacuum by
thermo- or photoactivated desorption of contaminants. The sample could be
thermostabilized to within 1 K over the range of 300$\div $700 K (or at
liquid nitrogen boiling point) in vacuum or in an atmosphere of admitted gas
at 10$^{-8}<$ P$<$10$^{-1}$ Torr. The photoelectron spectrometer chamber
contained a fused silica window trough which the specimen could be
illuminated with a UV-VIS mercury lamp at power density of 0.05 W/cm$^2$ in
the region of $h\nu $ $<$5.3 eV.

The energy diagram and the parameters determined in our experiments are
shown in Fig. 1. The value of the band gap of porous Si may vary from sample
to sample. Measurements in a diode configuration have given E$_g$ = 2.2 eV
[11] but larger values were obtained from electron energy loss spectroscopy
measurements (2.9 eV) [9] and by spectroscopic scanning tunneling microscopy
(2.6 eV) [12]. The Fermi level is located at the midgap region [9]. The
displacement of the photoelectron spectrum as a whole means the change of
the band bending V$_S$. The value of the work function $\varphi _T$ is
obtained from the equation $\varphi _T=h\nu -E_{kinmax}$ , where $E_{kinmax}$
is defined directly in the experiment. The shift of the highest binding
energy edge of the spectrum (for which E$_{kin}$= 0) means a change of the
dipole component $\delta $ of the work function $\varphi _T$ by $\Delta
\varphi _T$ = $\Delta $V$_S$ + $\Delta \delta $ . Thus it is possible to
determine the evolution of the electrophysical parameters $\varphi _T$, $%
\delta $, and $\Delta $V$_S$ during treatment of the surface. The shape of
the spectrum gives the DOS structure.

As can be concluded from the photoelectron spectra in Figs. 2 and 3 the
strongest feature is the A band with the maximum around 3.2 eV when the
specimen is heated in UHV at 600 K. After additional heating in the oxygen
atmosphere of p = 0.5 Torr the band shows a rigid shift of 0.25 eV toward E$%
_F$. This shift is due to appearing or increasing of the Shottky barrier V$_S
$ (Fig. 2, curves 1,2). The increasing of V$_S$ and change of $\delta $ mean
that (i) oxygen adsorbtion takes place and (ii) the electrons are
transferred from the volume of the sample to the adsorbing molecules,
forming negatively charged species as O$_2^{-}$ and O$^{-}$. The high
temperature adsorption of oxygen decreases the intensity of the minor band B
while the intensity of the major band A is unchanged (compare curves 1 and
2, Fig. 2). This suggests that the states responsible for B are surface
states while the A band is due to bulk-like states. The same effect may be
observed due to illumination of the sample (wavelengths longer than 185 nm)
in oxygen atmosphere at room temperature (Fig. 2, curve 3). This means that
photoactivated adsorption of oxygen takes place. Note that the red
luminescence drops drastically after thermo- or phototreatment in oxygen
atmosphere. In contrast to this behaviour the short- wavelength luminescence
retains its original value.

A sharp reduction in the intensity of the red photoluminescence band of
porous Si has been observed after annealing in UHV conditions ( 1$\cdot $10$%
^{-7}$ Torr) at temperatures $>$ 300 $^{\circ }$C [13,14]. This effect can
be ascribed to dangling bonds (trapping centers for nonradiative
recombination) created by desorption of hydrogen from the SiH$_2$ surface
species. Also UV illumination with an energy exeeding 3 eV may cause
hydrogen loss from porous Si in a nonoxidizing ambient [15]. In addition
thermally and optically induced oxidation is known to rapidly degrade the
photoluminescence intensity [16]. This coinsides with results of electron
paramagnetic resonance measurements [17] showing formation of disordered
dangling bond centers in porous Si annealed thermally in oxygen at
temperatures between 420$\div $770 K. We can assign the red luminescence
with process es induced by diffusion of photogenerated electrons from bulk
to the surface and subsequent recombination on the surface states. The
luminescence efficiency is weakened by (i) the Shottky barrier due to
charges localized on the thermo- or photoadsorbed oxygen layer and (ii) by
formation of dangling bond centers which activate nonradiative transition
processes. This conjecture is supported by the observation that the B band
of the photoelectron spectrum (see Fig. 2), exhibiting deeper states, is
typical to those parts of the specimen which inhibit the red photo-
luminescence.

It is noteworthy that some fine structure is observed on the low-energy tail
of the photoelectron spectra (see inserts to Figs. 2 and 3). The nature of
these states requires additional analysis. Thus in this paper we present for
the first time energy structure of porous silicon, described with
macroscopic ( $\varphi _T$, E$_F$, V$_S$, $\delta $) and microscopic (DOS)
parameters and their evolution during degradation of photoluminescence of
porous silicon.

Acknowledgments. This work was supported the Academy of Finland and the
Russian National Program ''Surface Atomic Structures'' (project No. 95-1.14).

{\em References }

1. L. T. Canham, Appl. Phys. Lett. 57, 1046 (1990). 

2. M. S. Brandt, H. D. Fuchs, M. Stuzmann, J. Weber, and M. Cardona, Solid
State Commun. 81, 307 (1992). 

3. F. Koch, V. Petrova-Koch, and T. Muschik, Proc. Symp. E on Light Emission
from Silicon, Strasbourg, France, May 4-7, 1993. 

4. J. C. Vial, A. Bsiesy, F. Gaspard, R. Hrino, M. Ligeon, F. Muller, R.
Romenstein, and R. M. Macfarlane, Phys. Rev. B 45, 14171 (1992). 

5. L. T. Canham, M. R. Houlton, W. Y. Leong, C. Pickering, and J. M. Keen, J.

Appl. Phys. 70, 422 (1991). 

6. K. Inoue, K. Meahashi, and H. Nakashima, Jpn. J. Appl. Phys. 32, L361
(1993). 

7. J. Terry, H. Liu, R. Cao, J. C. W oicik, P. Pianetta, X. Yang, J. Wu, M.
Richter, N. Maluf, F. Pease, A. Dillon, M. Robinson, and S. George, Chemical
Surface Preparation, Passivation and Cleaning for Semiconductor Growth and
Processing Symposium, San Francisco, USA, 27-29 April 1992 (Pittsburgh, PA,
USA: Mater. Res. Soc. 1992), p. 421. 

8. R. Guerrero-Lemus, J. L. G. Fierro, D. Moreno, and J. M. Martinez-Duart,
Mater. Sci. Technol. 11, 711 (1995). 

9. P. H. Hao, X. Y. Hou, F. L. Zhang, and Xun Wang, Appl. Phys. Lett. 64,
3602 (1994). 

10. A. M. Aprelev, A. A. Grazulis, A. I. Ionov, A. A. Lisachenko, and D. A.
Shuliatev, Phys. Low-Dim. Structures 10, 31 (1994). 

11. Zhiliang Chen, Tzung-Yin Lee, and Gijs Bosman, Appl. Phys. Lett. 64,
3446 (1994). 

12. R. Laiho and A. Pavlov, Phys. Rev. B 51, R14774 (1995). 

13. C. Tsai, K.-H. Li, J. Sarathy, J. C. Campbell, B. K. Hance, and J. M.
White, Appl. Phys. Lett. 56, 2814 (1991). 

14. S. M. Prokes, O. J. Glembocki, V. M. Bermudez, and R. Kaplan, Phys. Rev.
B 45, 13788 (1992). 

15. R. T. Collins, M. A. Tischler, and J. H. Stathis, Appl. Phys. Lett. 61,
1649 (1992). 

16. M. A. Tischler, R. T. Collins, J. H. Stathis, and J. C. Tsang, Appl.
Phys. Lett. 60, 639 (1992). 

17. R. Laiho, L. S. Vlasenko, M. M. Afanasiev, and M. P. Vlasenko, J. Appl.
Phys. 76, 4290 (1994).

\vspace{1.0in}{\em Figure captions.}

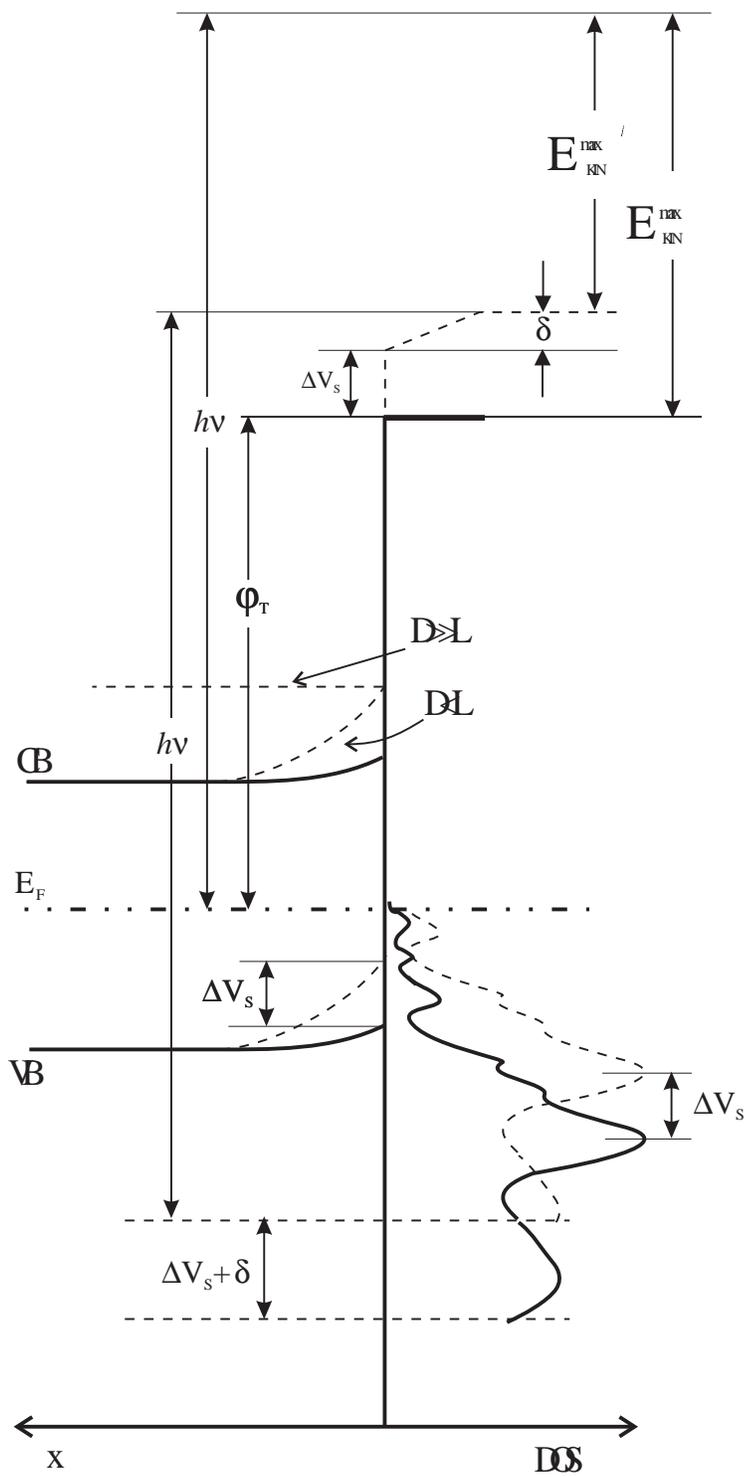
Fig. 1. Energy diagram of porous Si. $h\nu $ is the exitation energy (8.43
eV), $\varphi _T$ is the work function, $\delta $ is the dipole component of 
$\varphi _T$; CB and VB are the conduction and the valence bands edges, $%
\Delta $V$_S$ is an additional band bending due to sample tratments, D is
the Debay radius, L is the sample size, E$_{kin}$is the electron kinetic
energy in vacuum, $x$ is the geometrical coordinate from surface and N(DOS)
is the DOS magnitude.

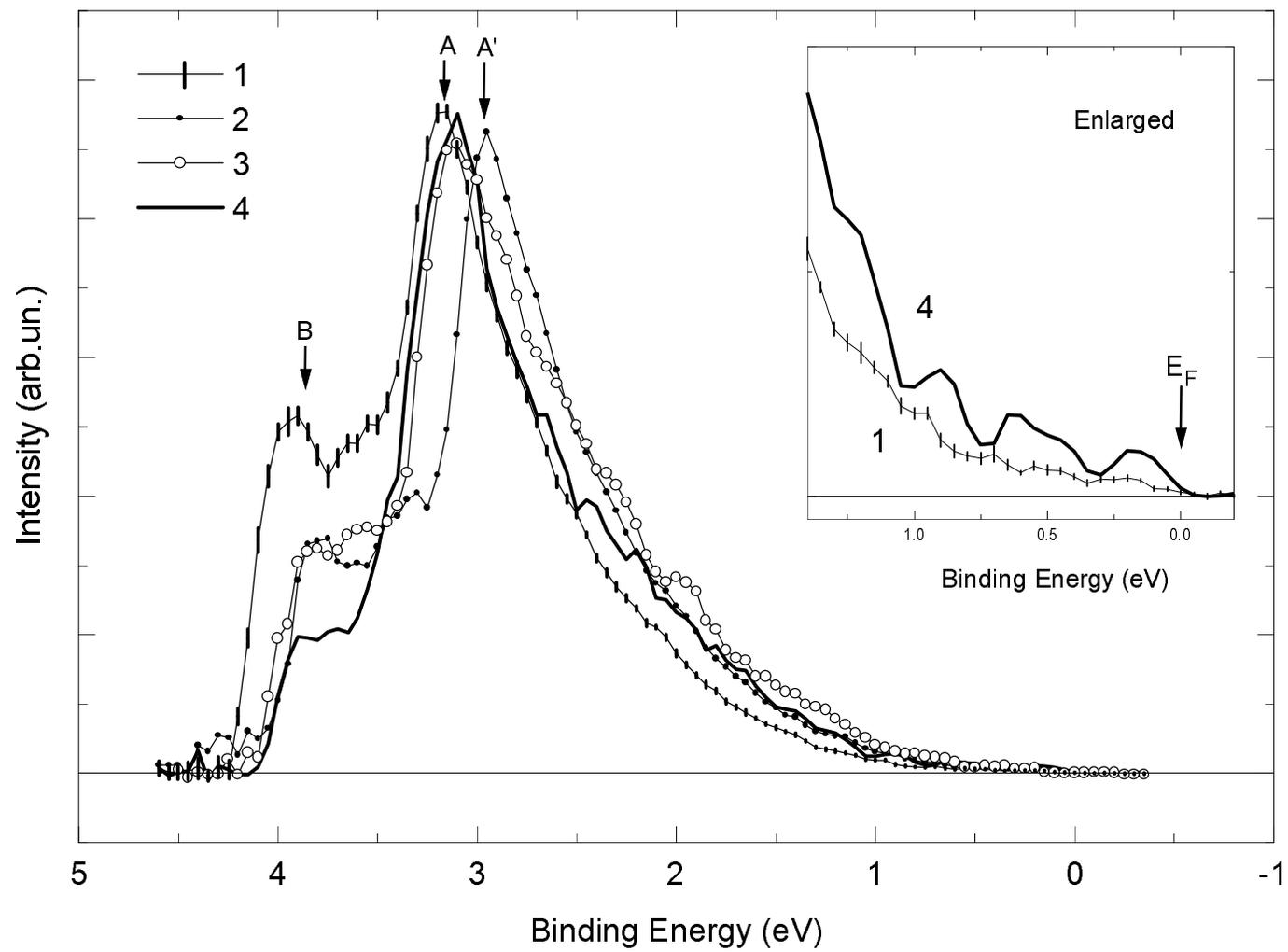
Fig. 2. Evolution of the photoelectron spectra of porous Si from the domains
of dominating yellow photoluminescence: (1) ''as received'' sample after
vacuum heating at 600 K, (2) after additional heating in O$_2$ at 0.5 Torr
at 600 K, (3) sample (1) after phototreatment in O$_2$ at 0.5 Torr, and (4)
sample (1) after heating in atmosphere at 600 K.

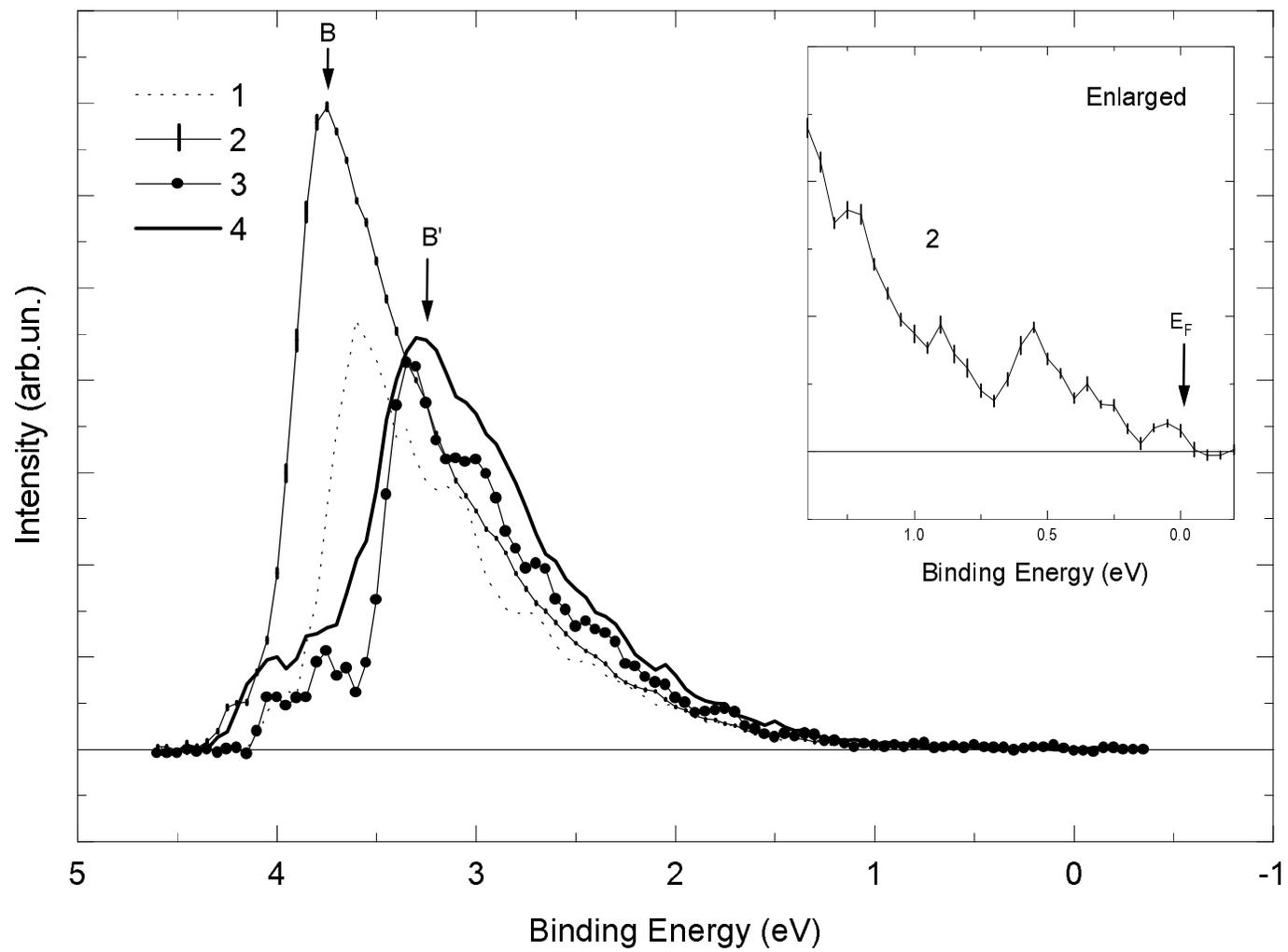
Fig. 3. Evolution of the photoelectron spectra of PS sample from the domains
of dominating red photoluminescence: (1) ''as received'' sample, (2) after
heating in high vacuum at 600K, (3) the same as (2) plus heating in O$_2$
(0.5 Torr) at 600 K and (4) the same as 3 plus heating in atmosphere at 600
K.

\end{document}